\documentclass[sigconf,nonacm]{acmart}

\renewcommand\footnotetextcopyrightpermission[1]{} 
\def\BibTeX{{\rm B\kern-.05em{\sc i\kern-.025em b}\kern-.08emT\kern-.1667em\lower.7ex\hbox{E}\kern-.125emX}}

\usepackage{multirow}
\usepackage{graphics}
\usepackage{amsmath}
\usepackage{mathrsfs}
\usepackage[caption=false]{subfig}
\usepackage{graphicx}
\usepackage{hhline}

\usepackage{url}
\usepackage{tikz}
\usetikzlibrary{shapes.geometric}
\usetikzlibrary{positioning}
\usetikzlibrary{arrows}

\usepackage[utf8]{inputenc}
\usepackage[english]{babel}
\usepackage{microtype}
\def\TN{\mathrm{TN}}
\def\TP{\mathrm{TP}}
\def\FN{\mathrm{FN}}
\def\FP{\mathrm{FP}}
\def\INC{\mathrm{INC}}

\newcommand{\stego}{\boldsymbol S} 
\newcommand{\pred}{\boldsymbol C}

\begin{document}

{\Huge
This is a preprint of the paper:\\

\noindent Daniel Lerch-Hostalot and David Meg\'{\i}as. 2023. ``Real-world actor-based image steganalysis via classifier inconsistency detection''. In \emph{Proceedings of the 18th International Conference on Availability, Reliability and Security (ARES '23).} Association for Computing Machinery, New York, NY, USA, Article 43, pp. 1–9. \hyperlink{https://doi.org/10.1145/3600160.3605042}{https://doi.org/10.1145/3600160.360504}
}
\newpage

\title{Real-world actor-based image steganalysis via classifier inconsistency detection} 

\author{Daniel Lerch-Hostalot \and David Meg\'{\i}as}
\affiliation{%
  \institution{Internet Interdisciplinary Institute (IN3), Universitat Oberta de Catalunya (UOC),\\ CYBERCAT-Center for Cybersecurity Research of Catalonia}
 \streetaddress{Rambla del Poblenou 154}
 \city{Barcelona}
 \country{Spain}
 \postcode{08018}
}
\email{{dlerch,dmegias}@uoc.edu}

\begin{abstract}
In this paper, we propose a robust method for detecting guilty actors in image steganography while effectively addressing the Cover Source Mismatch (CSM) problem, which arises when classifying images from one source using a classifier trained on images from another source. Designed for an actor-based scenario, our method combines the use of Detection of Classifier Inconsistencies (DCI) prediction with EfficientNet neural networks for feature extraction, and a Gradient Boosting Machine for the final classification. The proposed approach successfully determines whether an actor is innocent or guilty, or if they should be discarded due to excessive CSM. We show that the method remains reliable even in scenarios with high CSM, consistently achieving accuracy above 80\% and outperforming the baseline method. This novel approach contributes to the field of steganalysis by offering a practical and efficient solution for handling CSM and detecting guilty actors in real-world applications.

\end{abstract}



\maketitle

\section{Introduction}

Steganography involves various methods for concealing confidential information within seemingly harmless items. In contemporary times, these items primarily consist of digital media, with digital images serving as the most prevalent medium for steganographic purposes due to their extensive usage. Conversely, steganalysis encompasses an array of approaches employed to identify messages that were previously concealed through steganographic techniques.

The majority of contemporary steganalytic approaches employ machine learning techniques \cite{Fridrich:2012:RM, Denemark:2014:selchannel, Boroumand:2019:SRNet}. In steganalysis that adopts machine learning, an initial step consists of using a labeled dataset containing known cover and stego images to train a classification model. Subsequently, this trained model is applied to classify images in a testing dataset as either cover or stego.

This method performs exceptionally well under laboratory conditions, meaning that the training image set is akin to the testing image set employed by the steganographer to conceal confidential information. However, in real-world scenarios, the collection of media used by the steganographer might significantly differ from the training set used for the classifier \cite{Ker:2013:real_world}. This happens, for instance, when the testing set images are inadequately represented within the training set. Examples of this discrepancy include images captured with a distinct camera or resolution, images that have been compressed, magnified, or enhanced via filters, or images taken in considerably dissimilar conditions. In steganalysis, this issue is referred to as cover source mismatch (CSM) and was first documented in \cite{Cancelli:2008:csm}. Additional circumstances can result in imprecise predictions, such as stego source mismatch (SSM), which occurs when certain embedding parameters, like the precise payload \cite{Pevny:2011} or the stegosystem used \cite{butora:2019:ssm}, vary between the training and testing datasets.

Various methods have been proposed to address the CSM problem. In Ref. \cite{Fridrich:2014:csm_mitig}, three strategies were proposed: (1) Training with a combination of cover sources; (2) using different classifiers trained with various sources and testing with the closest one; and (3) adopting the second approach but testing each image individually using the nearest source. The "islet approach" \cite{Pasquet:2014} involves a pre-processing stage where images are organized into clusters and a steganalyzer is assigned to each group. \cite{Xu:2015:csm} introduces a methodology to construct a comprehensive and representative training set efficiently. In Ref. 
\cite{Lerch-Hostalot:2015}, 
an unsupervised steganalytic method is proposed that can circumvent the CSM problem without requiring a training set. 
\cite{Lerch-Hostalot:2019} 
proposes a technique to evaluate classifier performance and ensure accuracy in the context of CSM cases. 

The most usual approach to address the CSM problem involves constructing a comprehensive training dataset with images from various cameras and processing pipelines \cite{Xu:2015:csm, cogranne:2019:alaska:into_the_wild, cogranne:2020:alaska2,ruiz:2021:lssd}. This strategy aims to enable machine learning models to concentrate on universal features that allow for precise image classification across different sources. Nevertheless, a complete database that can effectively mitigate the CSM is yet to be developed.

An effective solution 
\cite{Lerch-Hostalot:2019,Megias:2023}
to the CSM problem involves attempting to detect it in order to prevent image misclassification. This strategy, which employs the DCI method 
\cite{Lerch-Hostalot:2019}, 
serves as the foundation for this article.

The remaining of this paper is organized as follows. In Section \ref{sec:preliminaries}, we discuss the context in which our proposed method operates and reflect on the relevance of this scenario for real-world applications. Section \ref{sec:background} introduces essential definitions and notations used to explain the Detection of Classifier Inconsistencies (DCI), which forms the foundation of our proposed method. Section \ref{sec:proposed} details the method we propose for addressing actor-based steganalysis in practical settings, as well as the description of the baseline that will be used in the experiments. Section \ref{sec:experimental} outlines the experiments conducted and presents the results obtained. Finally, Section \ref{sec:conclusion} offers a conclusion to the paper and identifies potential directions for future research.

\section{Preliminaries}
\label{sec:preliminaries}
This section summarizes the rationale behind this paper and the actor-based scenario that is assumed.

\subsection{Rationale}
The use of machine learning techniques in steganalysis generates a difficult problem to adddress: The CSM. This problem arises when attempting to classify images from one source using a classifier that has been trained on images from another source. When this occurs, if the source from which the images to be classified come from does not have its statistical properties well represented in the training set, a significant reduction in the classification results arises.

Here, we define the source as the statistical characteristics of a specific database that has undergone certain physical processes during the taking of the images and/or digital alterations during their processing. These characteristics are common among all the images in the database. Among the parameters that affect these statistical properties, we highlight the model of the camera used (due to the hardware involved in taking the image, with which it has been manufactured) or the processing pipeline (filters used, resizing algorithms, or other techniques).

Although different image databases have been used in steganalysis research, the use of databases such as Bossbase \cite{Bas:2011:boss} or Alaska \cite{cogranne:2019:alaska:into_the_wild} is common. Ideally, the larger and more diverse the training database, the better chance we have that the images we want to classify are well represented. For this reason, different attempts have been made to create sufficiently large and diverse databases, such as iStego100K \cite{yang:2019:IStego100K} or Large Scale Steganalysis Database (LSSD) \cite{ruiz:2021:lssd}. Attempts have also been made to use databases with millions of images from other research fields, such as the ImageNet \cite{Deng:2009:imagenet} database.

However, to date, no result has evidenced the existence of a sufficiently diverse image database to suppress the CSM. While achieving a database with such characteristics is doubtful, it remains essential to have a methodology capable of handling CSM in order to apply steganalysis effectively in the real world.

\subsection{Actor-based scenario}

The proposed method is specifically designed to operate in an actor-based scenario, where various actors use images and may or may not apply steganography to conceal messages. To clarify, we distinguish between guilty actors who apply steganography and innocent actors who do not. 
This kind of stegoanalysis is known in the literature as batch or pool steganalysis \cite{Ker:2006:batch}.

Within an actor-based scenario, it is common for actors to use multiple images. While the analysis of individual images may be straightforward, the challenge arises when dealing with cases of CSM. However, in this article, we show that, in an actor-based scenario where multiple images are used, the CSM problem can be addressed quite successfully. 

The proposed framework is based on a few reasonable assumptions. Firstly, we assume that each actor has a sufficient number of images, all of which exhibit the same distribution. (i.e., they come from the same source). Secondly, we assume that the steganography algorithms used are included in the list of analyzed steganography algorithms in our framework.

\section{Background}
\label{sec:background}

This Section summarizes the idea of using subsequent embedding \cite{Megias:2023} 
exploited in the method called ``Detection of Classifier Inconsistencies'' (DCI), first introduced in 
\cite{Lerch-Hostalot:2019}, 
and extended and analyzed from a theoretical perspective in \cite{Megias:2023}. 

The method proposed in this paper goes beyond the method DCI proposed in our previous works and makes it possible to face a more realistic scenario.

\subsection{Definitions and notation}
In this section, we provide the basic definitions that are needed for the description of the method:
\begin{itemize}
    \item \textbf{Images:} Let $X$ be an image of $w\times h$ pixels, i.e. $X=\left(x_{i,j}\right)$ for  $i=1,\dots,w$ and $j = 1, \dots, h$, where $w$ and $h$ are, respectively, the width and the height of the image. If an image does not have any message embedded into it, it is called a ``cover'' image. Grayscale uncompressed images are considered for notational simplicity, but the proposed method can be applied for color and compressed images.
    \item \textbf{Steganographic function and stego images:} Let $\stego$ be a steganographic embedding algorithm such that $X'={\stego}_p^K(X,m)$, where $X=\left(x_{i,j}\right)$ is the input image and $X'=\left(x'_{i,j}\right)$ is the output (stego) image. This function requires some parameters $p$ (such as the embedding bit rate), a secret stego key $K\in\mathbf K$, and a secret message $m\in\mathbf M$. $\mathbf K$ and $\mathbf M$ are, respectively, the key and the message spaces.
    \item \textbf{Domain of ``steganalyzable'' images:} Let $\mathcal I$ be the domain formed by the union of the domains of all cover ($\mathcal C$) and all stego ($\mathcal S$) images, or $\mathcal I=\mathcal C \cup \mathcal S$.
    \item \textbf{Subsequent embedding:} Given an image $X\in \mathcal C$, the corresponding stego ($X'$) and ``double stego'' ($X''$) images are defined as follows: $X'={\stego}_p^{K_1}(X,m_1)$, and $X''={\stego}_p^{K_2}(X',m_2)={\stego}_p^{K_2}\left({\stego}_p^{K_1}(X,m_1),m_2\right)$, where he second embedding is carried out using the same parameters $p$, but a different stego key and message. The set of ``double stego'' images is denoted by $\mathcal D$. 
    \item \textbf{Separability conditions:} To apply the method described in this paper, the following three conditions are required: (i) $\mathcal C \cap \mathcal S=\emptyset$ (standard in steganalysis), (ii) $\mathcal S \cap \mathcal D=\emptyset$, and (iii) $\mathcal C \cap \mathcal D=\emptyset$. As discussed in \cite{Megias:2023}, conditions (ii) and (iii) can be satisfied if there are enough features satisfying directionality (see below) in successive embeddings. 
    \item \textbf{Primary classifier:} Given a collection of images ${\mathsf A}=\{A_i\}\subset \mathcal{I}$, we define the primary classifier as a prediction function: ${\pred}^{\mathsf A}_{{\stego}_p}:\mathcal{I}\rightarrow\{0,1\}$, where 0 means that the image is classified (predicted) as cover and 1 means that the image is classified (predicted) as stego. This is the standard steganalysis problem that can be found in the literature.
    \item \textbf{Secondary classifier:} Now, we can define the secondary set as $\mathsf B=\{B_i=\stego_p^{K_i}(A_i,m_i)\}\subset \mathcal I'$, where $\mathcal I' = \mathcal S \cup \mathcal D$.  I.e., $\mathsf B$ is the set obtained after a new embedding process in all the images of $\mathsf A$. Cover images in $\mathsf A$ become stego images in $\mathsf B$ and stego images in $\mathsf A$ become ``double stego'' images in $\mathsf B$. If the sets of stego and ``double stego'' images are separable (see the separability conditions above), we can define a secondary classification problem as follows: $\pred^{\mathsf B}_{{\stego}_p}: \mathcal I'\rightarrow\{0,1\}$, where $\pred^{\mathsf B}_{{\stego}_p}(B_i)=0$ means that  $B_i$ is classified (or predicted) as stego and $\pred^{\mathsf B}_{{\stego}_p}(B_i)=1$ means that $B_i$ is classified or predicted as ``double stego''.
    \item \textbf{Feature-based classification:} Both the primary and the secondary classifiers are assumed to work with a set of features, either explicit or implicit.\footnote{In may Machine Learning classifiers, especially artificial neural networks (ANN), the features are obtained by the ANN itself without any explicit feature extraction process.} The features can be modeled as a function $\mathcal{F}$ that takes an image $Z$ as input and outputs a real-valued vector. For an image $Z=\left(z_{i,j}\right)\in\{\mathcal{I}\cup \mathcal{I}'\}$, we have $\mathcal F(Z)=V\in\mathbb R^D$, where $D$ is the dimension of the feature vector, or $\mathcal F:\mathcal \{I \cup I'\} \rightarrow \mathbb R^D$. $\mathcal F$ can be given as a collection of $D$ real-valued functions $\mathcal F(\cdot)=\left[f_1(\cdot),\dots,f_D(\cdot)\right]^{\mathrm T}$.   
    \item \textbf{Directional features:}  We first define the operator $\Delta$ as the variation of the feature $f_i$ after data embedding, i.e: $\Delta f_i(X)=f_i\left(\stego^{K_1}_p(X,m_1)\right)-f_i(X)$, for a cover image $X\in \mathcal C$. We define a second-order variation $\Delta^2$ in a similar way: $\Delta^2 f_i(X)=f_i\left(\stego^{K_2}_p(\stego^{K_1}_p(X,m_1),m_2)\right)-f_i(\stego^{K_1}_p(X,m_1))$. Now, a feature $f_i$ is defined as directional if $f_i(X)\times \Delta^2 f_i(X)>0$, i.e., if the changes in the feature $f_i$ occur in the same direction after one and two data embeddings.
\end{itemize}

\begin{figure}[ht]
  \centering
  \includegraphics[width=.9\columnwidth]{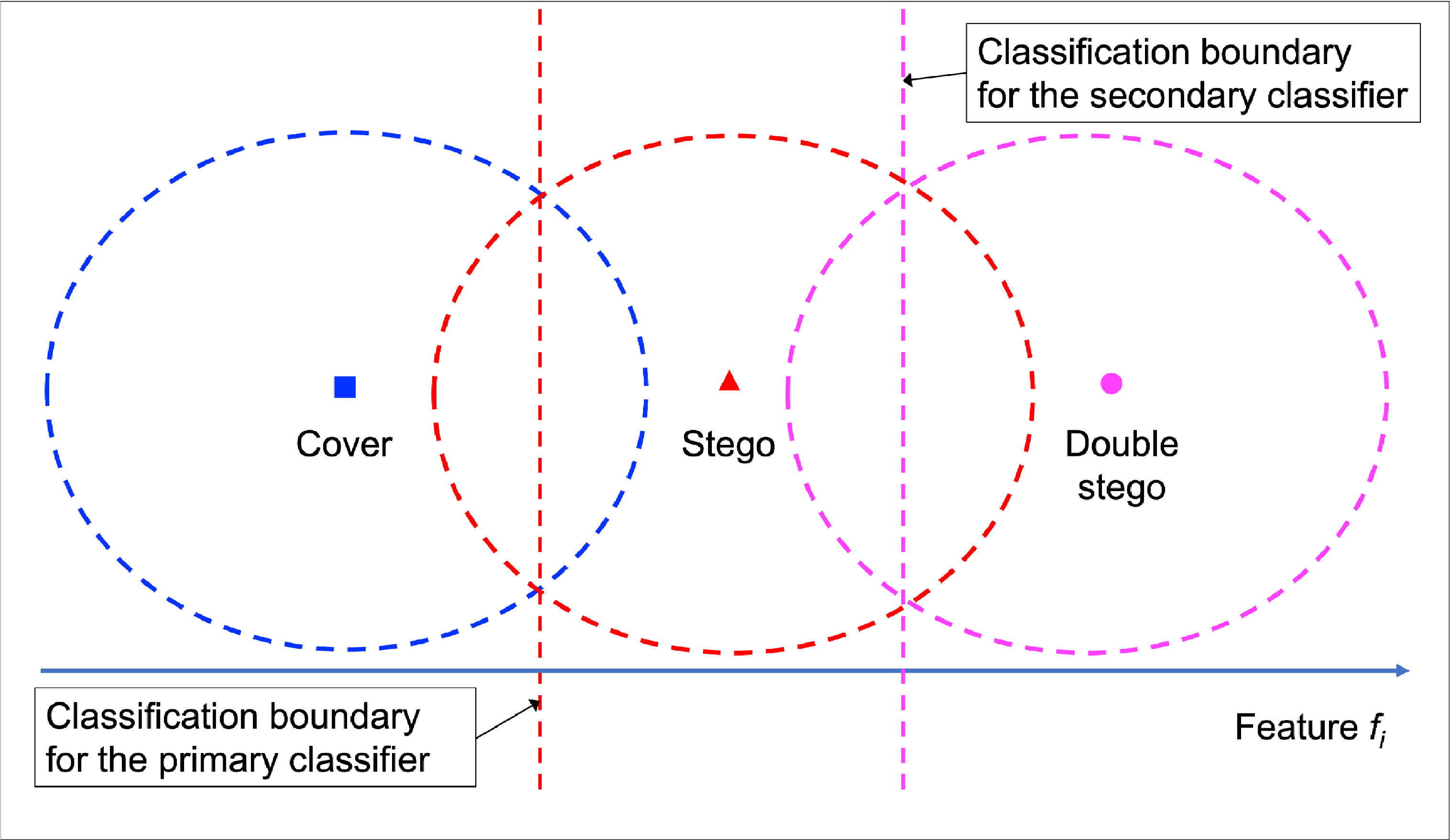}
  \caption{Sketch of a directional feature $f_i$.}
  \label{fig:directional}
\end{figure}

Fig. \ref{fig:directional} illustrates the concept of directional features. The abcissae represent the values of the feature $f_i$.  The centroids (average) values of the feature are represented with a square, a triangle and a circle for cover, stego and ``double stego'' images, respectively, whereas the dotted circles represent the ranges of the values of the feature for the three types of images. As it can be observed in the figure, the values of this feature for cover image are smaller (left side) than those of stego images (middle). Similarly, the values of $f_i$ for ``double stego'' images are greater than those of stego images. Hence, both $\Delta f_i$ and $\Delta^2 f_i$ are positive, and the feature $f_i$ is, thus, directional, according to the definition provided above.

\subsection{Detection of classification inconsistencies}
Fig. \ref{fig:directional} shows that the separability conditions can be achieved when there are (enough) directional features for a significant number of images. The primary and secondary classifiers can be trained to learn the boundaries between the different sets. Of course, in real conditions, some overlap will occur and the classification will never be perfect.

The primary and secondary classifiers can now be used to detect inaccurate classification of images, that may occur as a result of cover source mismatch, stego source mismatch, unknown message length or some other source of uncertainty 
\cite{Lerch-Hostalot:2019, Megias:2023}:

\begin{enumerate}
    \item We have a training set of images $\mathsf{A^{L}}=\{A_1,\dots A_{N_\mathsf L}\}\subset\mathcal{I}$. Without loss of generality, we assume that the first few images in $\mathsf{A^{L}}$ are cover and the remaining ones are stego, i.e. $\{A_1,\dots,A_{M_{\mathsf L}}\}\subset\mathcal{C}$ and $\{A_{M_{\mathsf L}+1},\dots,A_{N_{\mathsf L}}\}\subset\mathcal{S}$, with $1\leq M_{\mathsf L}\leq N_{\mathsf L}$. The training set $A^{\mathsf L}$ is used to train the primary classifier $\pred^\mathsf{A}_{{\stego}_p}$.

 \item We create the secondary training set by carrying out a subsequent embedding in all the images of $\mathsf{A^L}$, as $\mathsf{B^L}=\{B_1,\dots B_{N_\mathsf L}\}$ such that $B_i=\stego_p^{K_i}(A_i,m_i)$.
    Hence, it is clear that $\{B_1,\dots,B_{M_{\mathsf L}}\} \subset \mathcal{S}$ and $\{B_{M_{\mathsf L}+1},\dots,B_{N_{\mathsf L}}\}\subset\mathcal{D}$. This secondary training set is used to train the secondary classifier $\pred^\mathsf{B}_{{\stego}_p}$.

  \item The objective of steganalysis is to classify a testing set $\mathsf{A^T}=\{A'_1,\dots,A'_{N_{\mathsf T}}\}\subset\mathcal{I}$. The result of this classification is two sets of indexes, namely (i) $I_0=\{i_1,\dots,i_{M_{\mathsf T}}\}$ for predicted cover images: $\pred^\mathsf{A}_{{\stego}_p}(A'_j)=0$  for $j=i_1,\dots,i_{M_{\mathsf T}}$; and (ii) $I_1=\{i_{m_{\mathsf T}+1},\dots,i_{N_{\mathsf T}}\}$ for predicted stego images: $\pred^\mathsf{A}_{{\stego}_p}(A'_k)=1$  for $k=i_{M_{\mathsf T}+1},\dots,i_{N_{\mathsf T}}$, with $1\leq M_{\mathsf T} \leq N_{\mathsf T}$.
  
  \item We create the secondary testing set with a subsequent embedding as $\mathsf{B^T}=\{B'_1\dots,B'_{N_{\mathsf T}}\}$, with $B'_i=\stego^{K'_i}_p(A'_i,m'_i)$. Then, we classify $\mathsf B^T$ using $\pred^{\mathsf B}_{{\stego}_p}$, obtaining the indexes $I'_0=\{i'_1,\dots,i'_{M'_{\mathsf T}}\}$ for images predicted as stego and $I'_1 = \{i'_{M'_{\mathsf T}+1},\dots,i'_{N_{\mathsf T}}\}$ for images predicted as ``double stego'', with $1\leq M'_{\mathsf T} \leq N_{\mathsf T}$.
  
  \item \textbf{Consistency Filter (Type 1):} Each index $1,\dots,N_{\mathsf T}$ now belongs to either $I_0$ or $I_1$, and to either $I'_0$ or $I'_1$. If some index $i\in I_0$, this means that $A'_i$ is classified as cover by $\pred^{\mathsf A}_{{\stego}_p}$. Consequently, $B'_i=\stego^{K'_i}_p(A'_i,m'_i)$ should be classified as stego (not as ``double stego'') by $\pred^{\mathsf B}_{{\stego}_p}$, and we expect $i\in I'_0$. Conversely, if $i\in I_1$, we expect $i\in I'_1$. Then, images that are consistently classified by both classifiers are in $(I_0 \cap I'_0)\cup (I_1 \cap I'_1)$. On the other hand, images that are not consistently classified by the primary and secondary classifiers are the ones with indices in $I_{\mathsf{NC},1}=(I_0 - I'_0) \cup (I_1 - I'_1)$.

   \item \textbf{Consistency Filter (Type 2):} If there are enough directional features in the model $\mathcal F$, on classifying the images in $\mathsf A^{\mathsf T}$ using the secondary classifier, all of them should be classified as stego or $\pred^{\mathsf B}_{{\stego}_p}(A'_i)=0$, for all $i=1,\dots,N_{\mathcal T}$. The indices that do not satisfy this condition are kept in the set $I''_1=\{i''_{M''_{\mathsf T}+1},\dots,i''_{N_{\mathsf T}}\}$. This means that $\pred^{\mathsf B}_{{\stego}_p}(A'_i)=1$ for $i\in I''_1$. Similarly, if we classify the images of $\mathsf{B^T}$ using the primary classifier $\pred^{\mathsf A}_{{\stego}_p}$, with enough directional features, we do not expect any of the images be classified as cover, since all of them are stego or ``double stego''. Hence, we can collect another set of indexes with an inconsistent classification result: $I'''_0=\{i'''_1,\dots,i'''_{M'''_{\mathsf T}}\}$ such that $\pred^{\mathsf A}_{{\stego}_p}(B'_i)=0$, for $i\in I'''_0$. Both Type 2 inconsistencies can be combined in a single set as $I_{\mathsf{NC},2}=I''_1 \cup  I'''_0$.

   \item The images $A'_i$ with indices $i\in I_{\mathsf{NC},1} \cup I_{\mathsf{NC},2}$ would rather remain not classified (NC) since they lead to an inconsistent result among the four possible classification results: $\pred^{\mathsf B}_{{\stego}_p}(A'_i)$, $\pred^{\mathsf A}_{{\stego}_p}(B'_i)$, $\pred^{\mathsf A}_{{\stego}_p}(A'_i)$, $\pred^{\mathsf B}_{{\stego}_p}(B'_i)$. 
\end{enumerate}

\begin{table}[ht]
    \caption{Classification cases and final output. The ``Type'' column refers to the type of Consistency Filter that applies in that case, if any.}
    \label{tab:cases}
    \centering
    \begin{tabular}{||c|c|c|c||c|c||}
    \hhline{|t:====:t:==:t|}
 $\pred^{\mathsf B}_{{\stego}_p}(A'_i)$ & $\pred^{\mathsf A}_{{\stego}_p}(B'_i)$ &
 $\pred^{\mathsf A}_{{\stego}_p}(A'_i)$ & $\pred^{\mathsf B}_{{\stego}_p}(B'_i)$ &
Type & Output  \\  \hhline{|:====::==:|}
0 & \textcolor{black}{$0^*$} & 0|1 & 0|1 & 2 & NC \\ \hhline{|:====::==:|}
\textbf{0} & \textbf{1} & \textbf{0} & \textbf{0} & \textbf{---} & \textbf{Cover} \\ \hhline{||----||--||}
0 & 1 & \textcolor{black}{$0^*$} & \textcolor{black}{$1^*$} & 1 & NC \\ \hhline{||----||--||}
0 & 1 & \textcolor{black}{$1^*$} & \textcolor{black}{$0^*$} & 1 & NC \\ \hhline{||----||--||}
\textbf{0} & \textbf{1} & \textbf{1}& \textbf{1} & \textbf{---} & \textbf{Stego} \\ \hhline{|:====::==:|}
\textcolor{black}{$1^*$} & \textcolor{black}{$0^*$} & 0|1 & 0|1 & 2 & NC \\ \hhline{|:====::==:|}
\textcolor{black}{$1^*$} & 1 & 0|1 & 0|1 & 2 & NC \\ \hhline{|b:====:b:==:b|}
    \end{tabular}
\end{table}

Table \ref{tab:cases} shows the 16 different cases that can arise by considering the binary results (0 or 1) obtained by both classifiers applied to the images $A'_i$ and $B'_i$. Only two of the possible cases are consistent and lead to a classification of an image as cover or stego. The remaining 14 cases lead to an inconsistent result. In fact, even more than one possible inconsistency filter can apply at the same time. The values causing the inconsistencies are highlighted using a ``*'' sign. Note that some rows of the table represent a compressed version of four cases, when the results is an inconsistency irrespective of the outputs of $\pred^{\mathsf A}_{{\stego}_p}(A'_i)$ and $\pred^{\mathsf B}_{{\stego}_p}(B'_i)$.

This detection of inconsistencies has several practical applications, as shown in 
\cite{Lerch-Hostalot:2019,Megias:2023}. 

In particular, applications for predicting the classification error of standard steganalysis when CSM or SSM occurs, for estimating an unknown message length, are described in Ref. \cite{Megias:2023}. Some applications when even the embedding method is unknown are also discussed in that paper. In addition, subsequent embedding can also be used to obtain an unsupervised steganalysis approach, as described in \cite{Lerch-Hostalot:2015}.

One important step in this paper is to use DCI to predict the accuracy of the classifier for a given testing set (an actor in this study). To estimate the classification accuracy of the testing set, the number of inconsistencies can be used by computing the following expression 
\cite{Lerch-Hostalot:2019}:

\begin{equation}
\widehat{\mathrm{Acc}} = 1-\frac{\INC}{2\left|\mathsf{A^T}\right|},  
\label{eq:Acc}
\end{equation}
where $\widehat{\mathrm{Acc}}$ refers to the estimated accuracy of the primary classifier, $\INC$ stands for the total number of inconsistencies (determined by the classification results as shown in Table \ref{tab:cases}), and $\left|\cdot\right|$ represents the cardinality of a set. 

In fact, as shown in \cite{Megias:2023}, the previous equation applies when then number of stego and cover images in the testing set is the same. A slightly more complex equation can be obtained when the ratio of stego and cover images in the testing set differs from $0.5$. However, such a ratio will, in general, be unknown (although it can also be estimated). For this reason, the prediction of the accuracy of the classification for the case of a ratio of $0.5$ stego images is a good indicator of the performance of a standard steganalytic classifier.

\section{Proposed method}
\label{sec:proposed}

\subsection{Building blocks}
To apply the method proposed in this paper, two essential components are required. The first one is a model trained on each one of the stegosystems, which enables the application of the DCI method. The second one is an appropriate classifier designed for handling features.

For the first component, we make use of neural networks, specifically, the EfficientNet architecture, due to its demonstrated effectiveness in steganalysis \cite{Yousfi:2020:alaska2, Yousfi:2021:improv_effnet}. To apply the DCI method, two models trained on each of the stegosystems we aim to detect are required: One model for classifying  cover and stego images ($\pred^\mathsf{A}_{{\stego}_p}$), and another for differentiating between stego and double stego images ($\pred^\mathsf{B}_{{\stego}_p}$). Thus, we develop two distinct models for each of the stegosystems targeted for detection. These models are trained using payloads of varying sizes, selected at random. This approach enables classification without requiring knowledge of the specific payload employed by the steganalyst. Training the neural network in this manner has been shown to be effective and well-suited for DCI environments (see Section D.2 in Appendix D of the supplementary material\footnote{\url{https://ieeexplore.ieee.org/document/9722958/media}.} of \cite{Megias:2023}).

The second component is another classifier ($\pred^\mathsf{F}$) suitable for training and classifying using features. While multiple options exist for this step, we have opted for a Gradient Boosting Machine \cite{Friedman:2001:GBM}, a well-known algorithm that obtains good results in feature classification at a high speed. Nevertheless, it is important to note that other alternatives are available. This classifier is trained using the predictions generated by the aforementioned neural networks as features. Consequently, the classifier generates the final prediction, determining the innocence or guilt of an actor and identifying the specific stegosystem used in case of guilt. Finally, a threshold is employed to discard actors with a high CSM.

\subsection{Implementation details}
The objective of this method is to ascertain whether an actor is innocent or guilty of sending images with embeded steganographic content. Given that actors may possess a varying number of images, the proposed approach must take into account this variability. Consequently, we propose the following steps:
\begin{enumerate}
    \item For each of the stegosystems targeted for detection, the corresponding trained model is used to predict if images are cover or stego. This yields a feature representing the percent of stego images predicted for the actor. Another feature is the $\widehat{\mathrm{Acc}}$ itself. As a result, we obtain two features for each stegosystem we intend to detect.
    \item Leveraging the features obtained in the previous step, the Gradient Boosting Machine ($\pred^\mathsf{F}$) is trained  to predict which {stego\-sys\-tem} was used by a ``guilty'' actor, if any steganography were used at all. 
\end{enumerate}

In the final classification stage, the classifier $C^F$ determines the probabilities associated with the actor's innocence or the use of each of the analyzed $N$ stegosystems. The highest probability is then considered to determine if the actor is predicted as ``innocent'' or ``guilty''. If found guilty, this method enables us to identify the specific stegosystem employed by the actor. 

To select the corresponding $\widehat{\mathrm{Acc}}$ value, we rely on the previous prediction. For example, if the prediction suggests that the actor used the steganographic algorithm $A$, we select the $\widehat{\mathrm{Acc}}$ value associated with algorithm $A$ and check if it exceeds a predetermined threshold (e.g., $0.7$). If it does, we confirm the prediction. If not, the actor is labeled as CSM. When the prediction indicates innocence, we compute the average of all the $\widehat{\mathrm{Acc}}$ values from the models associated with all the stegosystems. This average is then used as the $\widehat{\mathrm{Acc}}$ value to determine if the actor should be labeled as CSM.

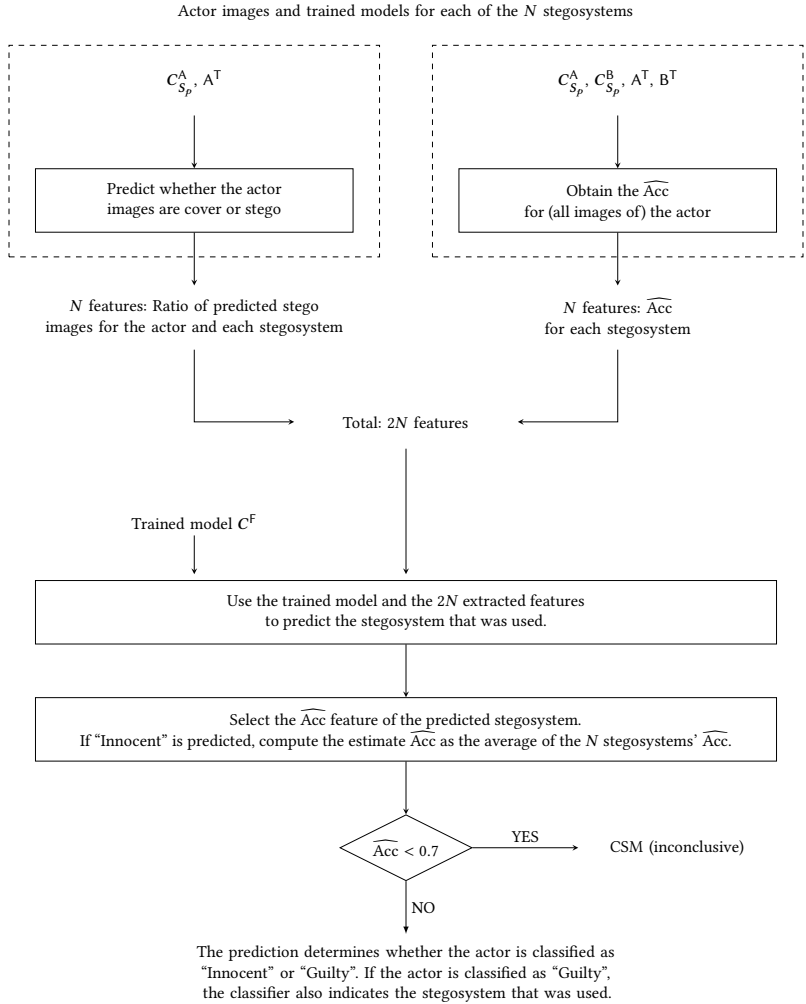
\begin{figure*}[ht]
\centering
\resizebox{0.6\textwidth}{!}{
\begin{tikzpicture} 


\node[text width=6cm, align=center] (L) at(-4cm,0cm){};
\node[text width=6cm, align=center] (C) at( 0cm,0cm){};
\node[text width=6cm, align=center] (R) at( 4cm,0cm){};



\node[text width=6cm, align=center, minimum width=4cm, minimum height=1.2cm, below=1cm of L] (input1)
{$\pred^{\mathsf A}_{{\stego}_p}$, $\mathsf{A^T}$};

\node[text width=6cm, align=center, minimum width=1cm, minimum height=1.2cm, below=1cm of R] (input2) 
{$\pred^{\mathsf A}_{{\stego}_p}$, $\pred^{\mathsf B}_{{\stego}_p}$, $\mathsf{A^T}$, $\mathsf{B^T}$};

\node[text width=14cm, align=center, below=0cm of C] (input)
{Actor images and trained models for each of the $N$ stegosystems};


\draw[dashed, dash pattern=on 3pt off 3pt] (-0.5,-1) rectangle (-7.5,-5);
\draw[dashed, dash pattern=on 3pt off 3pt] (7.5,-1) rectangle (0.5,-5);

\node[draw,
    align=center,
    minimum width=6cm,
    minimum height=1.2cm,
    below=1cm of input1
] (FE_LU) {Predict whether the actor\\ images are cover or stego};
 
\node[text width=6cm,
    align=center,
    minimum width=6cm,
    minimum height=1.2cm,
    below=1cm of FE_LU
] (FE_LD) 
{$N$ features: Ratio of predicted stego \\images for the actor and each stegosystem};

\draw[-stealth] (FE_LU.south) -- (FE_LD.north) node[midway,above]{};

\node[draw,
    align=center,
    minimum width=6cm,
    minimum height=1.2cm,
    below=1cm of input2
] (FE_RU) {Obtain the $\widehat{\mathrm{Acc}}$\\ for (all images of) the actor};

\node[text width=6cm,
    align=center,
    minimum width=6cm,
    minimum height=1.2cm,
    below=1cm of FE_RU
] (FE_RD) {$N$ features: $\widehat{\mathrm{Acc}}$\\ for each stegosystem};

\draw[-stealth] (FE_RU.south) -- (FE_RD.north) node[midway,above]{};

\draw[-stealth] (input1.south)-- (FE_LU.north) node[midway,above]{};
\draw[-stealth] (input2.south)-- (FE_RU.north) node[midway,above]{};

\node[text width=4cm, minimum height=1cm, align=center, below=7.5cm of C] (FE_output) 
{Total: $2N$ features};

\draw[-stealth] (FE_LD.south)|-(FE_output.west) node[midway,above]{};
\draw[-stealth] (FE_RD.south)|-(FE_output.east) node[midway,above]{};


\node[text width=4cm, align=center, below=3cm of FE_LD] (input3) {Trained model $\pred^\mathsf{F}$};


\node[draw,
    align=center,
    minimum width=14cm,
    minimum height=1.2cm,
    below=11cm of C
] (predict)
{Use the trained model and the $2N$ extracted features\\ to predict the stegosystem that was used.};

\draw[-stealth] (input3.south)--(-4,-11) node[midway,above]{};
\draw[-stealth] (FE_output.south)--(0,-11) node[midway,above]{};

\node[draw,
    align=center,
    minimum width=14cm,
    minimum height=1.2cm,
    below=1cm of predict
] (select)
{Select the $\widehat{\mathrm{Acc}}$ feature of the predicted stegosystem.\\ 
If ``Innocent'' is predicted, compute the estimate $\widehat{\mathrm{Acc}}$ as the average of the $N$ stegosystems' $\widehat{\mathrm{Acc}}$.};

\draw[-stealth] (predict.south)--(select.north) node[midway,above]{};

\node[draw, diamond, aspect=2, below=1cm of select](threshold) {$\widehat{\mathrm{Acc}} < 0.7$};

\draw[-stealth] (select.south)--(threshold.north) node[midway,above]{};

\node[text width=3.5cm,
    align=center,
    minimum width=1cm,
    minimum height=1.5cm,
    right=2cm of threshold
] (csm)
{CSM (inconclusive)};

\draw[-stealth] (threshold.east)--(csm.west) node[midway,above]{YES};

\node[
    text width=10cm, 
    align=center, 
    minimum height=1.5cm,
    below=1cm of threshold
] (final_output) 
{The prediction determines whether the actor is classified as ``Innocent'' or ``Guilty''. If the actor is classified as ``Guilty'', the classifier also indicates the stegosystem that was used.};

\draw[-latex] (threshold.south)--(final_output.north) node[midway,right]{NO};

\end{tikzpicture}
}
  \caption{Block diagram of the proposed scheme.}
  \label{fig:block_diagram}
\end{figure*}

Note that the threshold is a crucial parameter requiring fine-tuning. The  value chosen for this parameter, which can considerably impact the outcomes, is $0.7$. When the value of this threshold is increased, the method will yield more accurate results, albeit at the cost of a higher number of actors being labeled as CSM. 

A detailed block diagram of the proposed method is provided in Fig. \ref{fig:block_diagram}. In this figure, the standard steganalysis classification methodology is highlighted with a dashed rectangle in the top-left corner of the diagram. This consists of a classifier ($\pred_{S_p}^{\mathsf A}$) which predicts (actor) images either as cover (0) of stego (1). Using this ``standard'' steganalysis, we obtain a ratio of images predicted as stego for each analyzed stegosystem and each author. When this process is repeated for all $N$ stegosystems (i.e., $N$ different classifiers $\pred_{S_p}^{\mathsf A}$), $N$ ratios are collected and used as input features by another block below.

The other elements in the diagram are not typically found in conventional steganalys. First, highlighted with a dashed rectangle in the top-right corner, we find the DCI \cite{Lerch-Hostalot:2015,Megias:2023} predictor, which uses an additional classifier ($\pred_{S_p}^{\mathsf B}$) and the secondary testing set ($\mathsf{B^T}$) for each stegosystem. This block carries out three additional classification tasks for each image: $A'_i\in \mathsf{A^T}$ is classified with $\pred_{S_p}^{\mathsf B}$ as either stego (0) or ``double stego'' (1), and the corresponding embedded image $B'_i\in \mathsf{B^T}$ is classified with $\pred_{S_p}^{\mathsf A}$ as cover (0) or stego (1), and with $\pred_{S_p}^{\mathsf B}$ as stego (0) or ``double stego'' (1). Thus, for each image, we obtain four binary values that are combined using Table \ref{tab:cases}. Finally, taking into account all the inconsistencies for an actor, the estimated accuracy $\widehat{\mathrm{Acc}}$ is computed using Expression (\ref{eq:Acc}). Combining the predicted accuracies for all $N$ stegosystems, we obtain $N$ additional features to add to the $N$ ratios collected from the top-leftmost block. Hence, the total number of features per actor is $2N$.

The remaining blocks represent the final classification step carried out using the Gradient Boosting Machine mentioned above, which finally makes it possible to obtain a prediction of each actor as ``Guilty'', ``Innocent'', or inconclusive (CSM). The latter outcome occurs when the final estimated accuracy is lower than the selected threshold.

\subsection{Baseline (standard) method}
It is worth noting that the proposed method can be implemented without applying the DCI method, using only the information on the percent of stego images predicted by the model associated with each stegosystem as features. 

We employ this concept to develop a reference method, enabling us to compare the results obtained with our approach to those of a similar method that does not make use of DCI. This comparison makes it possible to assess the benefits of applying DCI within a steganalysis environment and to understand its effectiveness in addressing the CSM issue.

\section{Experimental results}
\label{sec:experimental}
This section presents several experiments that have been carried out to illustate the usefulness of the proposed approach.

\subsection{Experiment setup}
To conduct actor detection experiments, we must simulate a realistic scenario, where the most challenging issue faced in the real world is CSM. Consequently, we need to create a scenario incorporating CSM. To achieve this, we use image databases for training that significantly differ from those used in the testing sets, thereby introducing CSM.

It is important to note that including a portion of images from all the databases used in this study in the testing sets could potentially resolve CSM in our experiments. However, that is not our goal. Our aim is to ensure the testing set contains images with statistical properties that greatly differ from those in the training set, in order to simulate a CSM scenario akin to what we would encounter in the real world. This approach allows us to examine how our method would perform under realistic conditions. 

We have constructed two datasets for the experiments: One in the spatial domain and another in the JPEG transformed domain. Each dataset comprises multiple actors, including both innocent and guilty parties, some using images without CSM and others with CSM. Each actor uses between 10 and 50 images, and in the case of a guilty actor, the number of stego images ranges from 10\% to 100\% of the actor's total images. Parameters such as the percent of stego images, the steganography algorithm employed, and the payload used in each image have been randomly selected.

To ensure a realistic experiment, we employed both state-of-the-art and widely used stegosystems for the experiments in the spatial domain as well as the transformed domain. For the spatial domain, we used the state-of-the-art HILL \cite{Li:2014:hill} and UNIWARD \cite{Holub:2014:uniward} stegosystems, along with the popular embedding method, LSB matching, which is employed in numerous steganography tools such as OpenStego \cite{OpenStego}. Additionally, we incorporated SteganoGAN \cite{zhang:2019:steganogan}, a stegosystem based in  generative networks. For the transformed domain, we used the state-of-the-art  J-UNIWARD \cite{Holub:2014:uniward} and nsF5 \cite{Fridirch:2007:nsF5} stegosystems, as well as the widely-used Steghide \cite{Hetzl:2005:steghide} and Outguess \cite{Provos:2001:outguess} tools.

The prepared dataset consists of 10,000 actors for training and 4,000 actors for testing in the spatial domain, as well as 10,000 actors for training and 4,000 actors for testing in the JPEG transformed domain.

In the experiments we provide the results for a variable percent of actors with CSM. To create these datasets, we increase the number of actors with CSM by selecting them uniformly from the set of actors generated using images with CSM. At the same time, we reduce the number of actors without CSM in the dataset.

The neural network used for the classifiers $\pred^\mathsf{A}_{{\stego}_p}$ and $\pred^\mathsf{B}_{{\stego}_p}$ is an EfficientNet B0 architecture, implemented using the TensorFlow library \cite{abadi:2015:tensorflow}. Meanwhile, the final classifier $\pred^F$ was implemented using the Scikit-learn library \cite{pedregosa:2011:sklearn}, with default parameters. 

\subsection{Image databases}

The image databases used in the experiment are listed below. These databases come from various sources, ensuring an environment with CSM.

\begin{itemize}
\item The Alaska 2 database is the set of images from the Alaska 2 competition \cite{cogranne:2019:alaska:into_the_wild,cogranne:2020:alaska2}. This database is formed by 75,000 cover images taken with different cameras
and with different sizes and processing pipelines.

\item The BOSS database is the set of images from the Break Our Steganographic System! competition \cite{Bas:2011:boss}. This database comprises 10,000 cover images captured with seven distinct cameras. We used the raw version of the database, extracting color images and cropping a central 512x512 patch from each image.

\item The CAMID database is obtained from the IEEE Signal Processing Society's competition on Camera Model Identification \cite{stamm:2018:camid}. This database consists of 512x512 images in the spatial domain.

\item The MINI dataset (Mini-Imagenet \cite{Vinyals:mini}) is a dataset that contains 50000 JPEG images from Imagenet \cite{Deng:2009:imagenet}.

\item The FACES dataset \cite{Huang:faces} (Labeled Faces in the Wild) is a dataset that contains more than 13,000 JPEG images of faces. 

\cite{Huang:faces}

\end{itemize}

For training, the actors have been created using images from Alaska 2, while images from Alaska 2 and also from the other sources have been employed to generate the test actors. Since the training is created using Alaska 2 only, when test actors are obtained from databases other than Alaska 2, we consider them as actors with CSM.

\subsection{Results}
The results are provided in Tables \ref{tab:results_spatial} and \ref{tab:results_jpg}, where the accuracy obtained for the data sets in the spatial domain and in the transformed domain is shown. The reported accuracy results are given by the following equation:
$$ 
\mathrm{Accuracy} = \frac{\TP+\TN}{\TP+\FP+\TN+\FN},
$$
where $\TP$, $\TN$, $\FP$, and $\FN$ stand for the number of true positives, true negatives, false positives and false negatives, respectively, and the positive and negative results refer to the classification of actors as guilty (using steganography) or innocent (not using steganography). For example, a true positive is a guilty actor that is correctly classified as using steganography. 

In the tables, we can see the results for our baseline and for the proposed method. Each row shows the results for a certain percent of images with CSM, starting with 0\% CSM and progressively reaching 100\% CSM. This makes it possible to analyze how the proposed method evolves as CSM cases are added with respect to the baseline. Although we cannot be sure of what we would find in real-world scenarios, we can consider the cases with high CSM a good representation of what we would find in the real world.

The tables do not show the results that indicate the accuracy of the detection of the correct stegosystem. These results lack precision, as the techniques used for steganography tend to overlap. For instance, in the spatial domain, the HILL and UNIWARD methods use a form of LSB matching, which frequently leads to detection errors. There is also a degree of confusion between UNIWARD and HILL, as these methods differ only in the computation of the cost function. A similar situation arises with JPEG stegosystems. In any case, the objective of this actor-based scenario is to detect the use of steganography. The identification of the actual stegosystem used by an actor was not a goal of this research and is an open research issue to be considered in future works.

We can observe that, when the percents of CSM are low (0\%), the results obtained with the baseline and the proposed methods are quite similar. However, as we introduce more CSM cases, the accuracy of the proposed method remains high, while the baseline exhibits an increasing error rate. In situations with high percents of CSM (100\%), the baseline method becomes ineffective (with an accuracy in identifying guilty actors around 50\%, which means random guessing), whereas the proposed method continues to exhibit high performance. The proposed method does label a number of actors as ``actors with CSM'' and is unable to classify them. Although having unclassifiable actors is not ideal, it is still preferable to misclassification, which occurs with the baseline. Thus, we can employ the proposed method in real-world scenarios, while the baseline is not suitable for such applications.

In Table \ref{tab:results_spatial}, we observe the results for the experiment in the spatial domain. As shown, the baseline results become ineffective after 75\% of CSM, while the proposed method maintains its accuracy. However, starting from 75\%, we encounter 63\% unclassified actors, which increases to 84\% when CSM reaches 100\%.

\begin{table}[ht!]
\begin{center}
\caption{Accuracy of the classification of actors in the spatial domain for the baseline and the proposed methods.}
\label{tab:results_spatial}
\begin{tabular}{r|r|r|r}
\hline 
\textbf{\scriptsize CSM (\%)} & 
\textbf{\scriptsize Baseline} & 
\textbf{\scriptsize Proposed} & 
\textbf{\scriptsize Not classified} \\
\hline
  0\% & 1.00  &  1.00 & 0.0060 \\
 25\% & 0.88 & 0.99 & 0.2130 \\
 50\% & 0.75 & 0.98 & 0.4235 \\
 75\% & 0.62 & 0.95 & 0.6335 \\
100\% & 0.50  & 0.84 & 0.8450 \\
\hline
\end{tabular}
\end{center}
\end{table}

In Table \ref{tab:results_jpg}, we examine the results for the experiment in the transformed domain. Similar to the spatial domain, the baseline results lose their usefulness after 75\% of CSM, whereas the proposed method remains accurate. From 75\% CSM, we see 56\% unclassified actors, and this number rises to 67\% when CSM reaches 100\%.

\begin{table}[ht!]
\begin{center}
\caption{Accuracy of the classification of actors in the transformed domain (JPEG) for the baseline and the proposed methods.}
\label{tab:results_jpg}
\begin{tabular}{r|r|r|r}
\hline 
\textbf{\scriptsize CSM (\%)} & 
\textbf{\scriptsize Baseline} & 
\textbf{\scriptsize Proposed} & 
\textbf{\scriptsize Not classified} \\
\hline
  0\% & 1.00  & 0.99 & 0.1875 \\
 25\% & 0.89 & 0.97 & 0.3125 \\
 50\% & 0.78 & 0.94 & 0.4330 \\
 75\% & 0.67 & 0.90 & 0.5585 \\
100\% & 0.56 & 0.81& 0.6685 \\
\hline
\end{tabular}
\end{center}
\end{table}

As mentioned in the previous section, the threshold at which we consider an actor to have CSM is a parameter that requires adjustment. The tables display experiments using a threshold of 0.7. If we increase the threshold value, the method becomes more accurate in its results, but discards more actors due to CSM. Conversely, a lower threshold discards fewer actors, but leads to a decreased accuracy. For instance, if we use a threshold of $0.65$ in the spatial domain, we can reduce the percent of actors discarded by CSM from 84\% to 75\% (in the case of 100\% CSM). However, this adjustment results in a decrease in accuracy to 62\%.

\section{Conclusion}
\label{sec:conclusion}

In this paper, we present a method for detecting guilty actors in image steganography that is applicable in real-world scenarios, as it effectively handles CSM cases. The method can work with actors utilizing various stegosystems and a variable message length, hence effectively handling SSM as well.

The proposed method employs models based on EfficientNet neural networks to perform the DCI prediction, as well as an additional classifier that uses neural network predictions as input. This classifier identifies the stegosystem used, enabling it to leverage the DCI prediction to determine if an actor should be discarded due to excessive CSM.

We show that the method is suitable for real-world applications, as it can determine whether an actor is innocent or guilty, or discard them for having too much CSM. This is achievable in realistic cases, where each actor has only between 10 and 50 images, the stegosystem is unknown (although assumed to be one of those analyzed), the payload is unknown, and guilty actors may have some cover images.

The results reveal that even in scenarios with a high amount of CSM, the method remains reliable. We achieve an accuracy consistently above 80\%, even with 100\% CSM, and can effectively discard CSM cases that the baseline method classifies incorrectly.

For future work, it would be valuable to design a more extensive experiment, using a greater number of stegosystems and images from various sources. Additionally, it could be interesting to extend the method to other types of steganography, such as audio or video. Finally, the detection of the actual stegosystem used by an actor remains an open problem.

\begin{acks}
The authors acknowledge the funding obtained by the Detection of fake newS on SocIal MedIa pLAtfoRms (DISSIMILAR) project from the EIG CONCERT-Japan with grant PCI2020-120689-2 (Government of Spain), and to the   PID2021-125962OB-C31 ``SECURING'' project granted by the Spanish Ministry of Science and Innovation.

We gratefully acknowledge the support of NVIDIA Corporation with the donation of an NVIDIA TITAN Xp GPU card that has been used in this work.
\end{acks}

\bibliographystyle{ACM-Reference-Format}
\bibliography{main}

\end{document}